\documentclass[aps,pra,superscriptaddress,showpacs,notitlepage,twocolumn]{revtex4-1}
\usepackage{hyperref}
\usepackage{amsmath}
\usepackage{graphicx}
\usepackage{color}
\usepackage{braket}
\usepackage{verbatim}

\newcommand{\be}{\begin{equation}}
\newcommand{\ee}{\end{equation}}

\newcommand{\bs}{\boldsymbol}

\newcommand{\bk}{\bs{k}}

\renewcommand{\Im}{\mathop{\mathrm{Im}}\nolimits}
\renewcommand{\Re}{\mathop{\mathrm{Re}}\nolimits}

\newcommand{\pcsadd}{Center for Theoretical Physics of Complex Systems, Institute for Basic Science, Daejeon 34126, Korea}

\newcommand{\vincaadd}{Vin\v{c}a Institute of Nuclear Sciences, University of Belgrade, National Institute of the Republic of Serbia, P.O.B. 522, 11001 Belgrade, Serbia}

\newcommand{\addtuc}{School of Electrical and Computer Engineering, Technical University of Crete, Chania, Greece 73100}

\newcommand{\addcqt}{Centre for Quantum Technologies, National University of Singapore, 3 Science Drive 2, Singapore 117543}

\begin{document}


\title{Nonlinear signatures of Floquet band topology}

\author{\firstname{Aleksandra} \surname{Maluckov}}
\affiliation{\vincaadd}

\author{\firstname{Ekaterina} \surname{Smolina}}
\affiliation{Institute of Applied Physics, Russian Academy of Science, Nizhny Novgorod 603950, Russia}

\author{\firstname{Daniel} \surname{Leykam}}
\affiliation{\addcqt}

\author{\firstname{Sinan} \surname{G\"undo\u{g}du}}
\affiliation{\pcsadd}
\affiliation{Department of Physics, Humboldt-Universit\"{a}t zu Berlin, Newtonstrasse 15, 12489 Berlin, Germany}
\author{\firstname{Dimitris G.} \surname{Angelakis}}
\affiliation{\addcqt}
\affiliation{\addtuc}

\author{\firstname{Daria A.} \surname{Smirnova}}
\affiliation{Nonlinear Physics Centre, Australian National University, Canberra ACT 2601, Australia}
\affiliation{Institute of Applied Physics, Russian Academy of Science, Nizhny Novgorod 603950, Russia}

\date{\today}

\begin{abstract}
We study how the nonlinear propagation dynamics of bulk states may be used to distinguish topological phases of slowly-driven Floquet lattices. First, we show how instabilities of nonlinear Bloch waves may be used to populate Floquet bands and measure their Chern number via the emergence of nontrivial polarization textures in a similar manner to static (undriven) lattices. Second, we show how the nonlinear dynamics of non-stationary superposition states may be used to identify dynamical symmetry inversion points in the intra-cycle dynamics, thereby allowing anomalous Floquet phases to be distinguished from the trivial phase. The approaches may be readily implemented using light propagation in nonlinear waveguide arrays. 
\end{abstract}

\maketitle

\section{Introduction}

The great progress and open challenges in the world of topological matter have provoked investigation of the topological phenomena in periodically-driven quantum systems. The particularities of driven systems allow a number of interesting topological phenomena to occur, including quantized pumping and the existence of novel ``anomalous'' topological phases with no analogue in undriven static systems~\cite{goldman,kitagawa,Rudner2013,ann,driving,jedan}.

Topological phases of driven systems are classified according to the properties of their Floquet operators, i.e. the time-evolution operators describing the dynamics over one period of the drive~\cite{Floquet_review_1,Floquet_review_2}. Each eigenstate of the Floquet operator accumulates a phase $\phi$ over one driving period, giving the opportunity to define a 'quasi-energy' $\Omega=\phi/T$, the average phase accumulated per unit time. Being a phase variable, the quasi-energy is periodic with period $2\pi/T$ and thus Floquet systems lack well-defined ground states. Nevertheless, they can still support topological bulk-edge correspondences whereby the number of protected edge modes appearing in finite systems is given by topological invariants of the bulk Floquet eigenstates~\cite{Rudner2013, ann, jedan}.

 Photonic systems provide a highly flexible platform for implementing periodically-driven systems and observing their topological properties~\cite{topo_review1,topo_review2}. For example, in optical waveguide arrays the longitudinal propagation distance plays the role of time, allowing Floquet phenomena to be realizing using purely spatial modulation of the refractive index~\cite{zamajt,photonic_floquet,Leykam2016,AFI_1,AFI_2,Afzal2020}.

Nonlinear effects are now being considered in the context of topological photonic systems~\cite{NonlinearTopo_review}, opening new opportunities for measurement and manipulation of topological bands and giving rise to new classes of self-localized nonlinear modes~\cite{Lumer2013,Lumer2016,Leykam2016b}. For example, the nonlinear modulational instability of Bloch waves may be used to populate energy bands of two-dimensional undriven lattices and measure their bulk topological invariants~\cite{MI_review,topo_MI}. Recently, bulk and edge solitons in topological band gaps were experimentally observed using two-dimensional Floquet waveguide lattices~\cite{Floquet_soliton,edge_soliton_1,edge_soliton_2,nonlinear_pump}. These recent achievements directed our interests to how nonlinearity may be used to probe the bulk topological invariants of Floquet systems.

Here we would like to see how the predictions of our earlier study Ref.~\cite{topo_MI} may be applied to nonlinear Floquet topological systems such as optical waveguide arrays. These systems can be well-described by effective static Hamiltonians in the case of high-frequency modulation; in this limit the results of Ref.~\cite{topo_MI} are directly applicable. Therefore the focus of this work is the behaviour in the low frequency driving regime, which in particular can lead to the emergence of anomalous Floquet topological phases.

To this end, we analyze modulational instability in a stroboscopically-driven nonlinear square lattice model. Each cycle of the periodic drive is divided into four steps. During each step pairs of neighbouring waveguides are coupled. By tuning the coupling strength during each step different topological phases hosting protected edge or corner modes can be realized, including the trivial insulator (TI), Chern insulator (CI), anomalous Floquet insulator (AFI), and anomalos Floquet higher order topological insulator (AFHOTI)~\cite{chong}. Similar to static lattices, transitions between distinct topological phases require the gap between the two bulk Floquet bands to close.

We show using numerical simulations of the governing nonlinear Schr\"odinger equation how the modulational instability dynamics can identify Chern insulator phases using the field polarization measures considered in Refs.~\cite{polarization,topo_MI}. These measures however cannot distinguish anomalous Floquet phases from the trivial insulator phases, which requires knowledge of the full time evolution within one driving period. To address this limitation, we consider the nonlinear dynamics of non-stationary plane wave-like states formed by equal superpositions of both bands of the considered Floquet lattice. We show that the nonlinear dynamics can be used to track the Floquet bands' symmetry eigenvalues at the high symmetry Brillouin zone points and thereby distinguish the anomalous Floquet phases from the trivial phase~\cite{chong}.

The outline of the article is as follows: The model equations and band structures of Floquet systems are considered in Section~\ref{sec:model}. The modulation instability is investigated in Section~\ref{sec:MI}, where we demonstrate the ability of MI to generate a quasi-steady state with well-defined polarization field. Next, in Sec.~\ref{sec:intra} we show how the dynamics of superposition states may be used to identify anomalous Floquet phases. Sec.~\ref{sec:conclusion} concludes.

\section{Floquet Lattice Model}
\label{sec:model}

We consider nonlinear light propagation in the periodically driven bipartite square lattice model of Ref.~\cite{chong}. Previous studies have considered similar models~\cite{Rudner2013,Leykam2016,Leykam2016b,AFI_1,AFI_2,chong,zhu2021}. The propagation dynamics are described by the nonlinear Schr\"odinger equation
\be 
i \partial_z \ket{\psi_{\bs{r}} (z)} = \left[ \hat{H}_L(z) + \hat{H}_{NL}(\psi_{\bs{r}}) \right] \ket{\psi_{\bs{r}}}, \label{eq:nlse}
\ee
where the propagation distance $z$ plays the role of time, $\ket{\psi_{\bs{r}}} = (a_{\bs{r}},b_{\bs{r}})^T$ encodes the optical field amplitude on the two sublattices, $\hat{H}_{L}(z)$ is the periodically-modulated linear tight binding Hamiltonian, and $\hat{H}_{NL}(\psi_{\bs{r}}) = g \, \mathrm{diag}[f(|a_{\bs{r}}|^2),f(|b_{\bs{r}}|^2)]$ describes the on-site nonlinearity with strength $g$, which is a function of the local intensity at each lattice site. For linear stability analysis we consider pure Kerr nonlinearity $f(I) = I$, while for the beam propagation simulations we use saturable nonlinearity $f(I) = I/(1+I)$. The latter reduces to Kerr nonlinearity in the limit of low intensities.

We start by reviewing the main properties of linear Floquet systems. As $\hat{H}_L(z)$ is a periodically-modulated lattice Hamiltonian, it gives rise to Floquet-Bloch modes $\ket{u_n(\bs{k})} e^{i \bs{k}\cdot \bs{r}}$ with wavenumber $\bs{k}$, which are invariant (up to a phase factor) under spatial translations by a lattice vector and after one period of the modulation, i.e.
\begin{align}
\hat{U}_L(\bs{k}, Z) \ket{u_n(\bs{k})} &= \exp \left( -i \int_0^Z \hat{H}_L(\bs{k},z) dz \right) \ket{u_n (\bs{k})} \nonumber \\
&=  \lambda_n(\bs{k}) \ket{u_n (\bs{k})},
\end{align} 
where $\hat{U}_L(\bs{k})$ is the linear Floquet evolution operator, whose eigenvalues $\lambda_n(\bs{k})$ forms a discrete set of bands indexed by $n$. The Floquet evolution operator can be used to define a static effective Floquet Hamiltonian $\hat{H}_F$ via $\hat{U}_L(\bs{k}, Z) = \exp(-i Z \hat{H}_F(\bs{k}))$ analogous to Hamiltonians of non-driven lattices, with bands of quasi-energy eigenvalues $\Omega_n$ defined modulo $2\pi/Z$
\begin{equation}
\Omega_n(\bs{k})=\frac{i}{Z}\ln\lambda_n(\bs{k}).
\end{equation}
The quasi-energy bands enable topological invariants originally derived for static systems to describe the stroboscopic dynamics of Floquet systems. For example, Chern insulator phases can be identified by integrating the Berry curvature of the Bloch wave eigenstates over the Brillouin zone to obtain the Chern number~\cite{topo_review1, topo_review2}
\be 
C = \frac{i}{2\pi} \int_{BZ} \left[ \braket{\partial_{k_x}u | \partial_{k_y}u} - \braket{\partial_{k_y}u | \partial_{k_x} u}\right]  {d}^2 {\bs k}. \label{eq:chern_number}
\ee
The Chern number of a band counts the difference between the number of chiral edge states emerging from the top and bottom of the band.

The periodicity of the quasi-energy eigenvalues leads to new topological phenomena inaccessible in static systems. Since the quasi-energies are only defined modulo $2\pi/Z$, it is possible for chiral edge states to traverse all of the band gaps, including what would be the semi-infinite gaps of a static system, corresponding to an anomalous Floquet insulator phase with vanishing Chern number. The topological invariants describing anomalous Floquet phases cannot be obtained using just $\hat{H}_F$ or $\hat{U}_L(Z)$, but instead require analysis of the full evolution throughout one driving period~\cite{Floquet_review_1,Floquet_review_2},
\be
 \hat{U}(z)={\cal{T}} \exp \left(- i \int_0^z H(z')dz' \right),
\ee 
where $\cal{T}$ denotes time ordering. For example, the anomalous Floquet phase is identified by the winding number~\cite{Rudner2013}
\be 
W = \frac{1}{8\pi^2} \int_0^Z dz \int_{BZ} d^2\bs{k} \mathrm{Tr}( \hat{U}^{-1} \partial_t \hat{U} [\hat{U}^{-1} \partial_{k_x} \hat{U}, \hat{U}^{-1} \partial_{k_y} \hat{U}]). \label{eq:AFI}
\ee 
The intra-period evolution operator $\hat{U}(z)$ is generally not periodic in time, despite the Hamiltonian obeying $\hat{H}_L(z+Z)=\hat{H}_L(z)$. For bulk systems the family of Bloch evolution operators $\hat{U}(\bs{k},z)$ acting within the space of periodic Bloch functions can be written in the form
\begin{equation}
\hat{U}(\bs{k},z)=\sum_{n=1}^N \hat{P}_n(\bs{k},z)\,\exp({-i\varphi_n(\bs{k},z)}),
\end{equation}
where $\hat{P}_n$ is the projector onto the $n$-th eigenstate of $\hat{U}(\bs{k})$, $\exp({-i\varphi_n(\bs{k},z)})$ the corresponding eigenvalues, and ${\varphi_n(\bs{k},z)}$ are called the phase bands of the system~\cite{microsing,fleng}. The phase bands depend on time and at $z=Z$ they are identical to the Floquet bands, i.e. the quasi-energies. Anomalous Floquet phases may be identified by band crossings of the phase bands, indicating that the system cannot be adiabatically deformed into a static (undriven) system.

\subsection{Linear Floquet spectrum}

\begin{figure}
    \centering
    \includegraphics[width=\columnwidth]{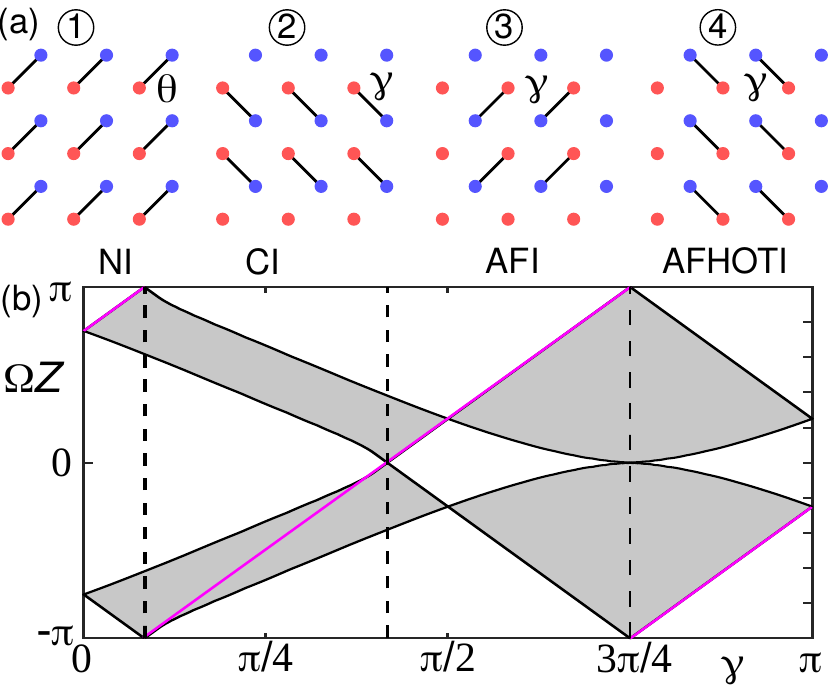}
    \caption{(a) Schematic of one driving period, consisting of each nonlinear site being coupled to its neighbours in turn with strength $J_{1}=\theta,\, J_{2-3}=\gamma$ for a duration $l=1$ (b) Linear Floquet-Bloch band structure as a function of $\gamma$ for fixed $\theta = 3\pi/4$. The purple line indicates the quasienergy of the symmetric Bloch wave with momentum $\bs{k}=0$. Dashed lines denote boundaries between different topological phases (normal insulator NI, Chern insulator CI, anomalous Floquet insulator AFI, and anomalous Floquet higher order topological insultor AFHOTI).}
    \label{fig:model}
\end{figure}

We now introduce and briefly review the properties of the linear lattice model ($g=0$). During each period $Z$ of the Floquet modulation each site is coupled once at a time to each of its four nearest neighbours with strength $J_j$, shown in Fig.~\ref{fig:model}(a) and described by the Bloch Hamiltonian
\be 
\hat{H}_L(\bs{k}, z) = \sum_{j=1}^4 \left(\begin{array}{cc} 0 & J_j(z) e^{i \bs{k}\cdot \bs{\delta}_i} \\  J_j(z) e^{-i \bs{k}\cdot \bs{\delta}_i} & 0 \end{array} \right),
\ee 
where $\bs{\delta}_1 = (\frac{1}{2},\frac{1}{2})$, $\bs{\delta}_2 = (\frac{1}{2},-\frac{1}{2})$, $\bs{\delta}_3 = -\bs{\delta}_1$, and $\bs{\delta}_4 = -\bs{\delta}_2$ are displacements between neighbouring sites. We normalize the lattice period to $Z=4$, and set $J_1(z) = \theta,\, J_2(z)=J_3(z)=J_4(z)=\gamma$ when the two neighbours are coupled, and zero otherwise~\cite{chong}.

The evolution operator can be written as
\be 
\hat{U}_L(\bk) =\hat{S}_{\gamma}(-\kappa_{-})\hat{S}_{\gamma}(-\kappa_{+})\hat{S}_{\gamma}(\kappa_{-})\hat{S}_{\theta}(\kappa_{+}),\label{Floquet_sup1}
\ee
where $\kappa_{+}=\bs{\delta}_1\cdot \bk, \,  \kappa_{-}=\bs{\delta}_2\cdot \bk$, and
\begin{eqnarray}
	\hat{S}_{J}(\kappa)= \left( \begin{array}{cc}  \cos{J} & -i e^{i \kappa}  \sin{J}\\ -i e^{-i \kappa} \sin{J} & \cos{J}\end{array} \right).
	\end{eqnarray}
Owing to our normalization of the coupling length to $Z/4 = 1$, $\hat{S}_{J}(\kappa)$ only depends on $J$ modulo $2\pi$, therefore we will use ``coupling strength'' and ``coupling angle'' interchangeably.

The quasi-energy band structure as a function of $\gamma$ for fixed $\theta = 3\pi/4$ is plotted in Fig.~\ref{fig:model}(b). Owing to the bipartite Hamiltonian which obeys particle-hole symmetry \cite{chong}, two quasi-energy gaps occur: around $\Omega=0$ and $\Omega=\pi/Z$. Each gap may host topological edge or corner modes. The gap closes at boundaries between distinct topological phases, corresponding to Dirac points in the bulk band structure.

The multiple crossing points in Fig.~\ref{fig:model}(b) indicate that several topological phases can be realized by tuning the single parameter $\gamma$: A normal insulator (NI) phase in which none of the gaps support any topological edge or corner modes; Chern insulator (CI) phases, in which only one gap hosts chiral edge modes; anomalous Floquet topological insulator (AFI) phases in which both gaps host chiral edge modes, and an anomalous Floquet higher-order topological insulator (AFHOTI) phase, in which both gaps host topological corner modes. 

The first Dirac point $\gamma=\pi/12$ corresponds to $\lambda_{1,2}=-1$ and divides the NI (Chern number $C=0$) and CI phases ($C=-1$). The second Dirac point $\gamma=5\pi/12$ corresponds to $\lambda_{1,2}=1$ and divides the CI and AFI phases, while the third Dirac point at $\gamma=3\pi/4$ and $\lambda_{1,2}=-1$ separates AFI and AFHOTI phases. In addition, there are other critical points at $\gamma = 0,  \pi/2$, and $\pi$, corresponding to purely flat quasi-energy bands with eigenvalues independent of the momentum $\bs{k}$. 

\begin{figure}
	\centering
	
	\includegraphics[width=\columnwidth]{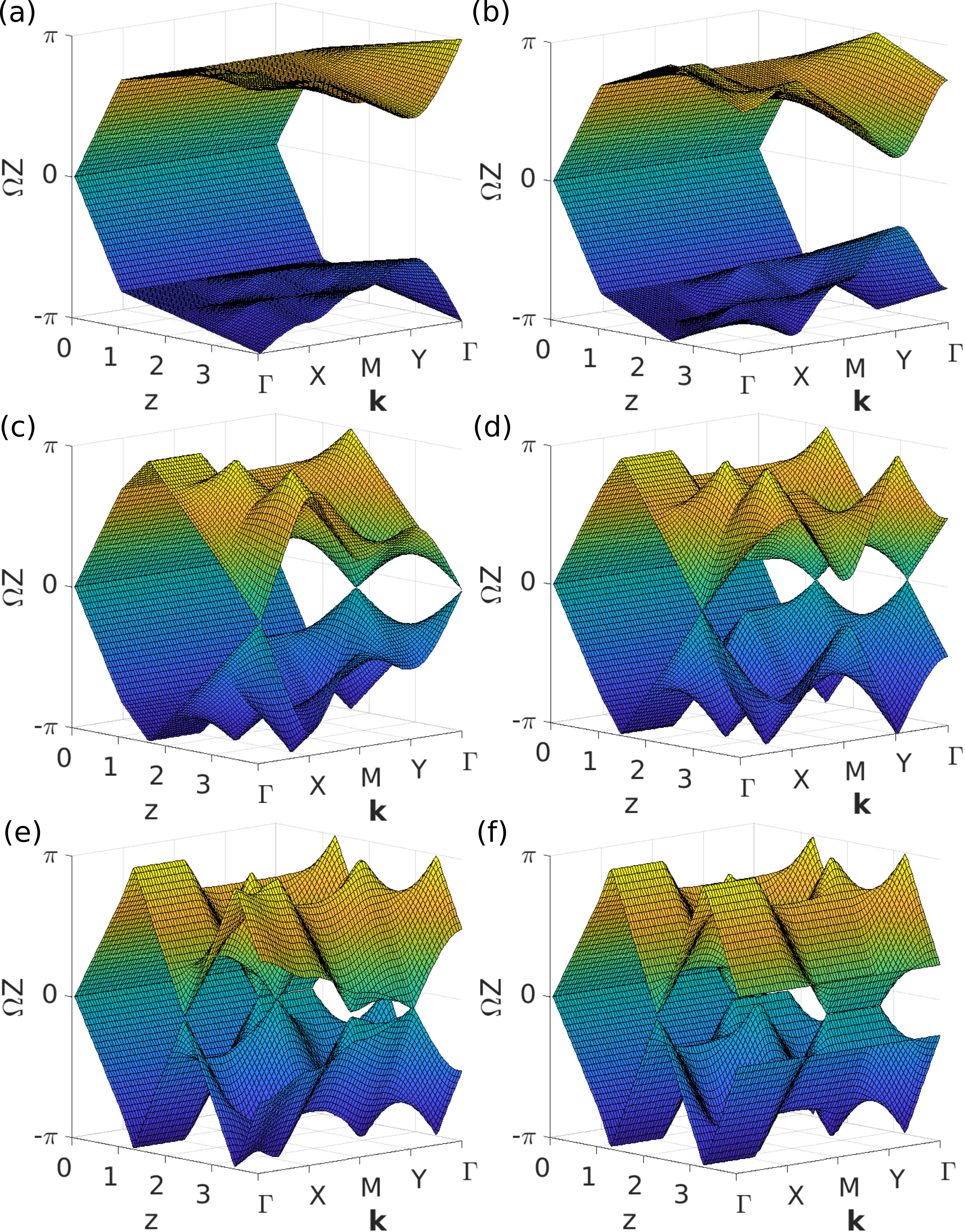}

	\caption{Phase band profiles along the high symmetry lines in the Brillouin zone for $\theta = 3\pi/4$ and (a) $\gamma = \pi/12$ (the first Dirac point), (b) $\gamma = \pi/6$ (Chern insulator phase), (c) $\gamma = 5\pi/12$ (the second Dirac point), (d) $\gamma = 7\pi/12$ (anomalous Floquet insulator phase), (e) $\gamma = 11\pi/12$ (anomalous Floquet higher order topological insualtor phase), and (f) $\gamma = \pi$ (flatband limit). Unremovable phase band crossings occur within the modulation cycle in the anomalous Floquet phases.}
	\label{fig:evs}
\end{figure}

Fig.~\ref{fig:evs} illustrates the phase band spectrum of the system for various $\gamma$. It is predicted and clearly seen in the corresponding phase band representation, that the phase band spectrum of the anomalous Floquet phases hosts degeneracies within the driving period. As degeneracies within a three-dimensional parameter space ($k_x,k_y,z$) they are topologically-protected; small changes in the lattice parameters can shift the positions of the degeneracies but not remove them. The presence of phase band degeneracies also means that the system cannot be adiabatically (without band crossing) transformed to an undriven system~\cite{microsing}. When viewing the lattice stroboscopically, i.e. only considering the quasi-energy bands, these degeneracies are hidden.

Recently, Ref.~\cite{chong} developed an alternative to Eq.~\eqref{eq:AFI} for obtaining bulk topological properties of the anomalous Floquet phases of this model. Because this model has chiral symmetry, it suffices to study the properties of the eigenstates at the high symmetry points of the Brillouin zone to distinguish the different phases. At the high symmetry points $[H(z),H(z^{\prime})] = 0$ for all $z, z^{\prime}$. Therefore the eigenstates of the phase operator $\hat{U}(z)$ are independent of $z$; all that changes are their eigenvalues. The number and position (quasienergy 0 or $\pi$) of the band crossings of $\hat{U}(z)$ at the high symmetry points can be used to distinguish trivial, Chern insulator, and anomalous Floquet topological phases.

\section{Modulational instability and dynamics}
\label{sec:MI}

Now we consider the properties of the full nonlinear evolution Eq.~\eqref{eq:nlse}. The linear Floquet modes may persist as nonlinear Floquet modes, solutions of the nonlinear eigenvalue problem
\begin{align}
\ket{u_{NL}(z+Z)} &= \hat{U}_{NL} \ket{u_{NL}(z)} \nonumber \\
&= \exp \left( -i \int_0^Z [\hat{H}_L(z) + \hat{H}_{NL}] dz \right) \ket{u_{NL}}\nonumber \\
&= e^{-i Z\Omega_{NL}} \ket{u_{NL}(z)}. \label{eq:NL_floquet}
\end{align}
As each step of the modulation cycle forms a nonlinear Schr\"odinger dimer, which is integrable, the dynamics are in principle analytically solvable. However, the solution will take a highly complicated form as the analytical solution of the dimer for arbitrary initial conditions involves elliptic functions~\cite{jensen}. Alternatively, Eq.~\eqref{eq:NL_floquet} can be solved numerically using the self-consistency method (see e.g. Refs.~\cite{Lumer2013,Leykam2016b}).

\begin{table} [t!]
\begin{center}
 \begin{tabular}{|c|c|c|c|c|c|c|c|c|c|c|}
\hline
Point & $\bs{k}_0$ & Eigenvalue & Eigenmode &
$\omega_1$& $\omega_2$& $\omega_3$& $\omega_4$\\\hline
$\Gamma$&(0,0) & $e^{-i(3\gamma+\theta)}$ & $(1,1)^{T}/\sqrt{2}$& 
$\theta$&$\gamma$&$\gamma$&$\gamma$\\\hline
$M$&($\pi$,$\pi$) &$e^{-i(\gamma-\theta)}$ & $(1,1)^{T}/\sqrt{2}$&
$-\theta$&$\gamma$&$-\gamma$&$\gamma$\\\hline
$X$&($\pi$,$0$) &$e^{i(\gamma-\theta)}$ & $(i,1)^{T}/\sqrt{2}$ & 
$\theta$&$\gamma$&$-\gamma$&$-\gamma$\\\hline
$Y$&($0$,$\pi$) &$e^{i(\gamma-\theta)}$ & $(i,1)^{T}/\sqrt{2}$ & 
$\theta$&$-\gamma$&$-\gamma$&$\gamma$\\\hline
\end{tabular}
\caption{Characteristics of the eigenstates  at the high-symmetry points. $\omega_j$ are linear eigenvalues of of $\hat{H}_L$ for the selected eigenvector at each quarter of the period. Eigenenergies of $\hat{H}_L$ are $\omega_j=\pm J_j$, and normalised eigenmodes are $|u_j (\bs{k}_0) \rangle=(\pm e^{i \bs{k}_0\cdot \boldsymbol{\delta}_j},1)^T/\sqrt{2}$.  
\label{tab:modes}}
\end{center}
\end{table}

In the following, we will restrict our analysis to the simplest nonlinear Bloch wave solutions, those at the high symmetry Brillouin zone points $\bs{k}_0$ (see Table~\ref{tab:modes}), which have modal profiles 
$|{u^{\pm}_{N\!L}(\bs{k}_0}) \rangle=(\pm e^{ i\Theta(\bs{k}_0)},1)^T/\sqrt{2}$,
independent of $\theta$, $\gamma$, and $g$, with unitary total intensity. Term $\Theta(\bs{k}_0)$ denotes the corresponding nonlinear mode phase. 
The normalised states $\ket{u_{\pm}} = (\pm 1,1)^T/\sqrt{2}$ at $\Gamma$ point have perhaps the simplest wave profile since no complex phase or intensity modulation is required, making these states more easily accessible in experiments~\cite{Floquet_soliton,edge_soliton_1,edge_soliton_2}.

There is no energy transfer between the sublattices and the Kerr nonlinear terms are independent of $z$, such that $\hat{H}_{N\!L} = g \hat{1}$, which commutes with $\hat{H}_L(z)$. Hence the nonlinear Bloch waves' quasienergies are $\Omega_{N\!L} = \Omega/Z + g /2$. Note that even if $g$ is small, the relative strength of the nonlinearity can be made increased by lengthening the modulation period $Z$. Because $\hat{H}_{NL}$ commutes with $\hat{H}_L$, these nonlinear Bloch wave eigenvectors are independent of $Z$.

We are interested in understanding how the features of the linear Floquet spectrum (e.g. dispersion, topological properties) are imprinted in the nonlinear Bloch waves' stability. Since $\hat{H}$ is invariant under the staggering transform, $(a_{\bs{r}},b_{\bs{r}})^T \rightarrow (-a_{\bs{r}},b_{\bs{r}})^T$, $z \rightarrow -z$, $g \rightarrow -g$, it is sufficient to consider only the symmetric Bloch waves $\ket{u_+} = (1, 1)^T/\sqrt{2}$. 

\subsection{Linear stability analysis of the nonlinear Bloch waves}

To enlighten the modulation instability (MI) development in the Floquet lattice, we study the stability properties of the nonlinear Bloch waves. Consider a small perturbation to a nonlinear Bloch wave,
$\ket{\psi_{\bs{r}}(z)}=\left(\ket{u^{+}_{N\!L}(\bs{k}_0)}+\ket{\delta \psi_{\bs{r}}} \right) e^{-i (\omega_{j} + g /2) z}e^{ i \boldsymbol{k}_0 \cdot \boldsymbol{r}}$, where $(\omega_j + g /2) $ is the nonlinear
mode's instantaneous energy at step $j$ of the modulation and $\boldsymbol{k}_0$ is the wave vector in the vicinity of which we investigate the instability.

We substitute this state into the evolution Eq.~\eqref{eq:nlse} for each of the modulation steps, retaining only the linear terms of the perturbation. To perform the linear stability analysis, we make use of the identities $\hat{\bs{k}} = - i \partial_{\boldsymbol{r}} $: $e^{i \hat{\bs{k}} \boldsymbol{\delta}_n} = \sum_n \dfrac{(i \hat{\bs{k}} \boldsymbol{\delta}_n)^n} {n!}= \sum_n \dfrac{( \partial_{\boldsymbol{r}} \boldsymbol{\delta}_n)^n} {n!}$, $e^{i \hat{\bs{k}} \boldsymbol{\delta}_n} [e^{i {\bs{k}} \boldsymbol{r}}]=e^{i {\bs{k}} \boldsymbol{\delta}_n} e^{i {\bs{k}} \boldsymbol{r}}$.

Considering perturbations of the form $\ket{\delta \psi_{\bs{r}}} = \ket{v_{\bs{p}}} e^{i \bs{p}\cdot\bs{r}} + \ket{w_{\bs{p}}^*} e^{-i \bs{p}\cdot\bs{r}}$ and applying the standard linear stability analysis, the Hamiltonian-like operator governing the evolution of the perturbation vector $\left(v,w,v^*,w^*\right)$ at each step of the driving period takes the form
\begin{widetext}
\be 
\hat{H'}_{j}=\left(\begin{array}{cccc}
g -\omega_{j} & J_{j} e^{i \bs{p}\cdot\bs{\delta}_{j}} e^{i \bs{k}_0\cdot\bs{\delta}_{j}} & \frac{1}{2} g e^{2i\Theta(\bs{k}_0)}& 0 \\
J_{j} e^{-i \bs{p}\cdot\bs{\delta}_{j}}e^{-i \bs{k}_0\bs{\delta}_{j}} & g-\omega_{j} & 0  & \frac{1}{2}g e^{2i\Theta(\bs{k}_0)} \\
-\frac{1}{2} ge^{-2i\Theta(\bs{k}_0)} & 0 & \omega_{j}- g & -J_{j} e^{i \bs{p}\cdot\bs{\delta}_{j}}e^{-i \bs{k}_0\cdot\bs{\delta}_{j}} \\
0 & -\frac{1}{2} ge^{-2i\Theta(\bs{k}_0)}  & -J_{j} e^{-i \bs{p}\cdot\bs{\delta}_{j}} e^{i \bs{k}_0\cdot\bs{\delta}_{j}} & \omega_{j}- g^2
\end{array}\right),
\ee
\end{widetext}
where $\omega_j=\pm J_j$ depends on the chosen mode (see Table~\ref{tab:modes}).

Using these Hamiltonians, we may construct the evolution operator for perturbations in the nonlinear case as
$\tilde{U}=e^{-i \hat{H_1}'}e^{-i \hat{H_2}'}e^{-i \hat{H_3}'}e^{-i \hat{H_4}'}$. If any of the eigenvalues $\tilde{\lambda}$ of $\tilde{U}$ have modulus $|\tilde{\lambda}|>1$ the nonlinear Bloch wave is unstable. 

The eigenvalues of $\tilde{\hat{H}}_{1,4}$ are $0,0,\pm 2 \sqrt{J_{1,4}(J_{1,4}-g)}$. Thus, a sufficient condition for instability is for one of the nonlinear couplers to be unstable, i.e. when $g > J_{1-4}$ ($J_1=\theta,\,J_{2-4}=\gamma$). This condition is independent of the lattice's band structure, and so less interesting from our point of view. Therefore we focus our attention to $g < J$ (i.e. $\gamma$ and $\kappa$). This is also the more practical case to consider, given the weakness of the Kerr nonlinear effect.

\begin{figure}
	\centering
			\includegraphics[width=\columnwidth]{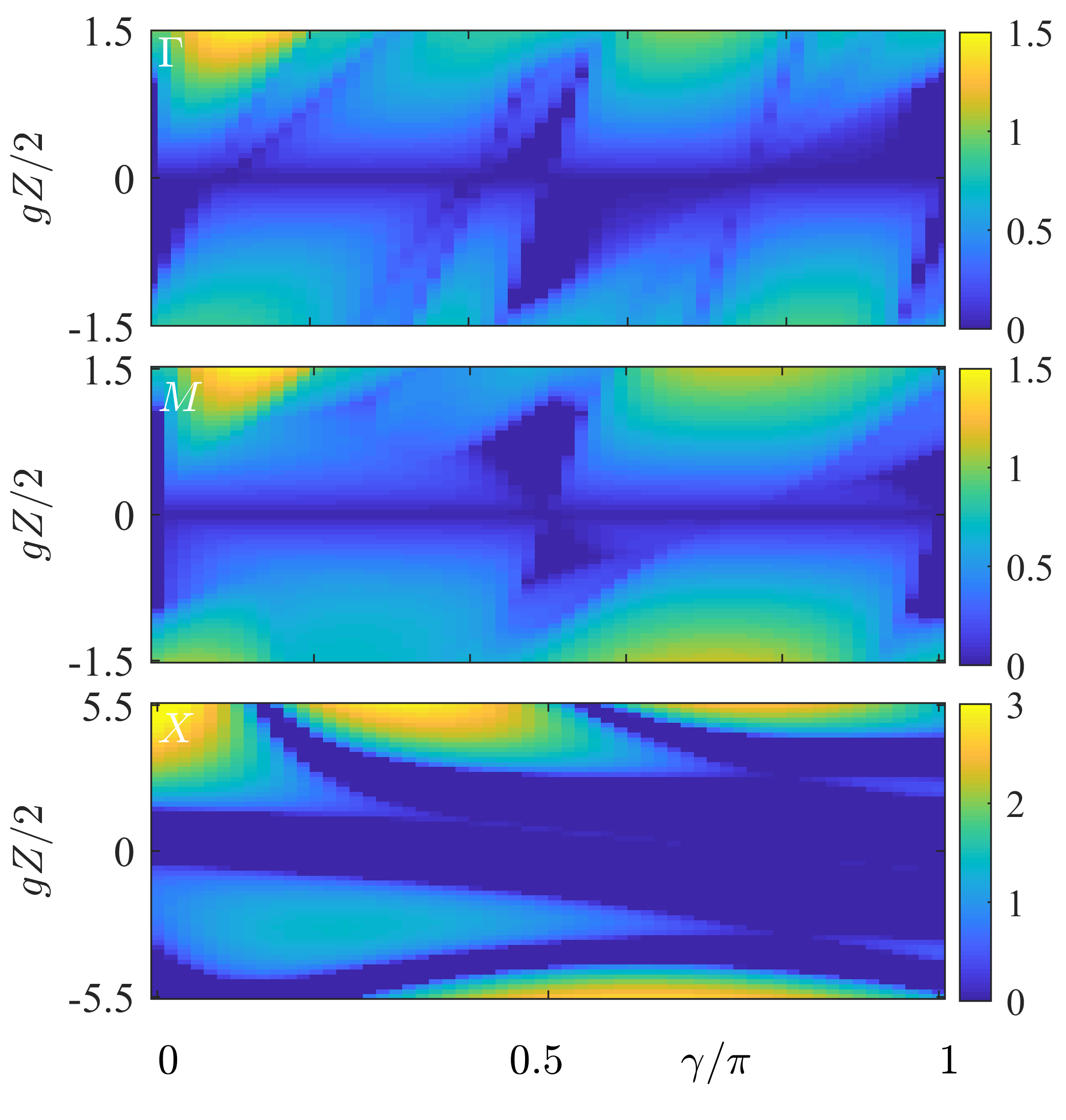} 
	\caption{The maximum values of the instability growth rate $\mathrm{max(\ln}|\lambda|)Z$ as a function of the coupling parameter $\gamma$ and nonlinearity strength $g$ for fixed $\theta = 3\pi/4$ at each of the high symmetry points of the Brillouin zone, labeled with $\Gamma$, $M$, $X$. 
	}
	\label{fig:tablsa}
\end{figure}

The numerically-obtained instability rates for the high symmetry points are shown in Fig.~\ref{fig:tablsa} as a function of $\gamma$ and $g$. For weak nonlinearity the stability windows are dictated by the quasi-energy of the linear Bloch wave. When the quasi-energy lies at a upper (lower) band edge, instability will occur for weak positive (negative) $g$. Otherwise, the nonlinearity needs to be sufficiently strong to shift the quasi-energy into a band gap for instabilities to occur. 

At the critical points of the band structure, i.e. the Dirac points and flatband limits, the sign of the wave effective mass flips, resulting in (for fixed $g$) transitions between stability and instability. These transitions can be seen for the $\Gamma$ and $M$ point Bloch waves, which lie at band edges. On the other hand, the $X$ and $Y$ points are typically saddles of the quasi-energy bands, meaning that sufficiently large nonlinearity is always required for the instability to develop. 

This behaviour of the linear stability eigenvalues closely resembles that of the Bloch waves of undriven lattices, see e.g. Ref.~\cite{topo_MI}. This is despite the modulation frequency being low and the presence of anomalous Floquet phases.

\subsection{Propagation dynamics}

Next we consider numerical simulations of the propagation dynamics to study the modulational instability beyond the linear perturbation approximation. We take as the initial state the $\bs{k}=0$ Bloch wave, $\ket{\psi_{\bs{r}}} = (1,1)^{T}/\sqrt{2}$ perturbed with noise (random perturbation amplitude of $10\%$), and solve the nonlinear Schr\"odinger equation with saturable nonlinearity using the split step method and periodic boundary conditions. We average over $100$ realizations of the initial random perturbation. To depict the findings of dynamical simulations in this section we have chosen the nonlinearity parameter $|g| = 0.35$ for which the initial state is linearly unstable in the three phases considered, with $g$ negative in the CI and AFHOTI phases and positive in the AFI phase.

\begin{figure}
	\centering
	\includegraphics[width=\columnwidth]{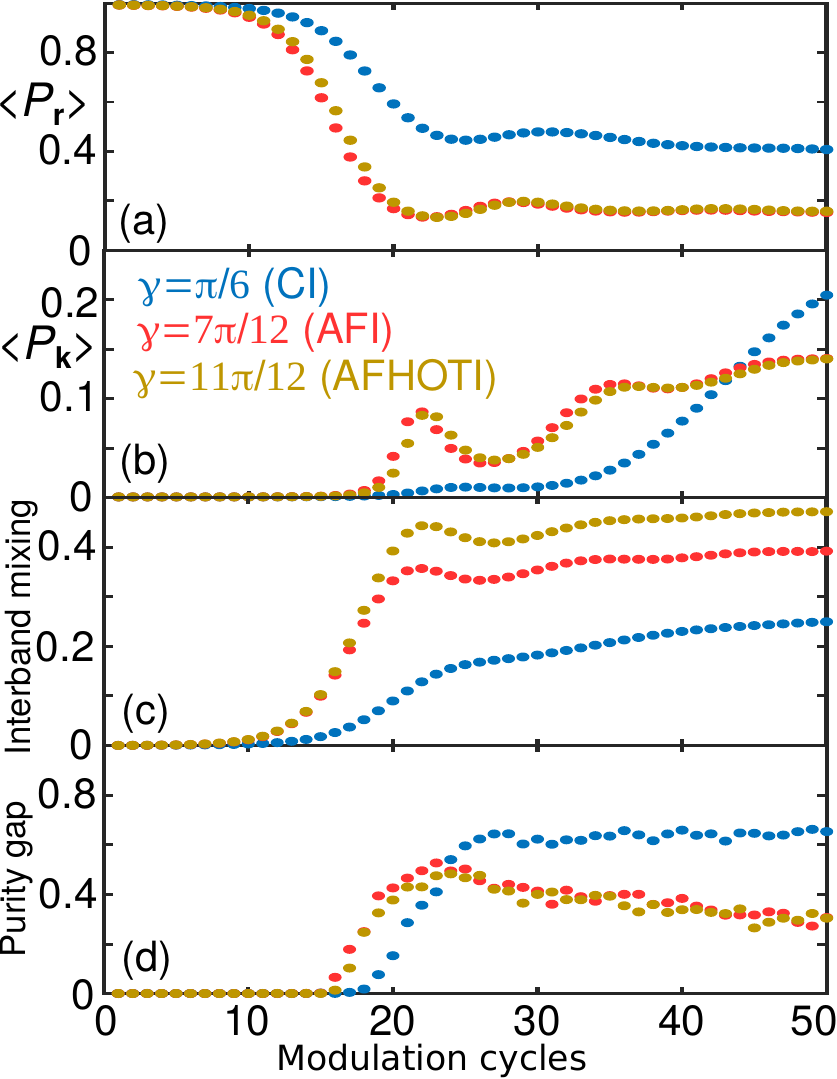}
	\caption{Examples of MI instability dynamics in the nontrivial Floquet phases. (a) The real space participation number, indicating self-focusing from the initial uniform beam. (b) The Fourier space participation number, indicating spreading in $\bs{k}$ space. (c) The fraction of energy transferred from the initially-excited band, indicating  that more energy remains in the initial band. (d) The purity gap, indicating the emergence of a well-defined polarization state throughout the entire Brillouin zone.} 
	\label{fig:MI_dynamics}
\end{figure}

To examine the propagation mode dynamics we follow the evolution of the normalized real space participation number $P_{\textrm{r}}$, the normalized Fourier space participation number $P_{\textrm{k}}$, the interband mixing strength, and the purity gap, whose dynamics are shown in Fig. \ref{fig:MI_dynamics}.

The real space participation number $P_{\textrm{r}}(z)$ determines the fraction of strongly excited lattice sites during propagation. It is defined by expression \cite{topo_MI} 
\be 
P_{\textrm{r}} = \frac{\mathcal{P}^2}{2N} \left( \sum_{\bs{r}}  |a_{\bs{r}}|^4 +  |b_{\bs{r}}|^4 \right)^{-1},
\ee
where $\mathcal{P} = \sum_{\bs{r}} \braket{\psi(\bs{r})|\psi (\bs{r})}$ is the total mode power and $N = 32 \times 32$ is the number of unit cells in the lattice, such that $0 \leq P_{\textrm{r}} \leq 1$.

The Fourier space participation number $P_{\textrm{k}}$ is the $k$-space equivalent of $P_{\textrm{r}}$,
\be 
P_{\textrm{k}} = \frac{\mathcal{P}^2}{2N} \left( \sum_{\bs{k}}  |a_{\bs{k}}|^4 +  |b_{\bs{k}}|^4 \right)^{-1},
\ee
which measures the fraction of excited Fourier modes. Averages of $P_{\textrm{r}}(z)$ and $P_{\textrm{k}}(z)$ over the ensemble of initial conditions are plotted in Fig.~\ref{fig:MI_dynamics}(a) and (b) respectively, for the CI, AFI, and AFHOTI phases.

To obtain the information on the efficiency of band mixing triggered by nonlinearity we compute the fraction of energy transferred from the initially-excited band to the rest of system. This is carried out by projecting the field profile at the end of each modulation cycle onto the basis of linear modes of the Floquet evolution operator $\hat{U}_L(\bs{k}, Z)$. When this fraction remains less than 0.5 the system retains some memory of the Floquet band that was excited at $z=0$. This is the case for all three phases considered, as shown in Fig.~\ref{fig:MI_dynamics}(c).

The extraction of information on the band topological properties requires an observable which imprints the band vector properties. Following Refs.~\cite{Hu2016,Bardyn2013,Budich2015,topo_MI} we consider the ensemble-averaged field spin (or polarization) textures $\bf{s}_{\psi}({\bs k})=\bra{\psi({\bs{k}})}\hat{\bs{ \sigma}}\ket{\psi({\bs{k}})}$, where $\hat{\bs{\sigma}} = (\hat{\sigma}_x,\hat{\sigma}_y,\hat{\sigma}_z)$ are Pauli matrices acting on the sublattice degree of freedom:
\begin{eqnarray}
s_{\psi\:x}({\bs k})&=&2\Re(a_{\bs{k}}^*b_{\bs{k}})/(|a_{\bs{k}}|^2+|b_{\bs{k}}|^2),\nonumber\\
s_{\psi\:y}({\bs k})&=&2\Im(a_{\bs{k}}^*b_{\bs{k}})/(|a_{\bs{k}}|^2+|b_{\bs{k}}|^2),\\
s_{\psi\:z}({\bs k})&=&(|a_{\bs{k}}|^2-|b_{\bs{k}}|^2)/(|a_{\bs{k}}|^2+|b_{\bs{k}}|^2).\nonumber\label{spintexture}
\end{eqnarray}

By averaging the field polarization over the ensemble of random perturbations added to the initially injected mode in the lattice we obtain the average polarization, $\langle\hat{\bf s}_{\psi}({\bs k})\rangle$, which is an experimentally measurable quantity. Mixed polarization states have $s_{\psi}^2 = \langle\hat{\bf s}_{\psi}({\bs k})\rangle \cdot \langle\hat{\bf s}_{\psi}({\bs k})\rangle < 1$.

Geometrically, the vector field $\langle s_{\psi}(\bs{k}) \rangle$ can be parameterized in terms of a polarization azimuth $\xi({\bs k})$ and polarization ellipticity $\chi({\bs k})$, which can be related to the amplitudes of the spinor components via~\cite{polarization}
\begin{align}
		\xi({\bs k})&=\frac{1}{2}\mathrm{tan}^{-1}\left(\frac{s_{\psi\:x}({\bs k})}{s_{\psi\:z}({\bs k})}\right)=\frac{1}{2}\mathrm{tan}^{-1}\left(\frac{2\mathrm{Re}(a^*_{\bs{k}}b_{\bs{k}})}{|a_{\bs{k}}|^2-|b_{\bs{k}}|^2} \right)\\
	\chi({\bs k})&=\frac{1}{2}\mathrm{sin}^{-1}\left(\frac{2\mathrm{Im}(a^*_{\bs{k}}b_{\bs{k}})}{|a_{\bs{k}}|^2+|b_{\bs{k}}|^2}\right).
	\label{polarization}	
\end{align}
Provided $s^2_{\psi}>0$ over the entire Brillouin zone, the polarization state is well-defined for all $\bs{k}$ and is sensitive to the Chern number Eq.~\eqref{eq:chern_number}. To be precise, the Chern number can be obtained as a sum of the phase singularities of the polarization azimuth $\xi$ weighted by $\mathrm{sgn}(s_{\psi\:y})$ at the singularity~\cite{polarization}. The Chern number computed using the polarization state is quantized and cannot change unless the polarization becomes undefined at some wavevector $\bs{k}_c$, i.e. $s^2_{\psi}(\bs{k}_c) = 0$. The purity gap $\mathrm{min}_{\bs{k}} (s_{\psi}^2)$ thus plays a role analogous to the band gap of the linear lattice. Fig.~\ref{fig:MI_dynamics}(d) illustrates the dynamics of the purity gap, illustrating the emergence of a nonzero purity gap via the modulational instability in all three phases.

This, even though the system is periodically driven (non-equilibrium), the modulational instability is able to generate a quasi-steady-state in which $P_{\textrm{r}}$, the band populations, and the purity gap converge to (approximately) time-independent values. On the other hand, $P_{\bs{k}}$ maintains a slow growth indicating that some memory of the initial state remains and an equilibrium state has not been reached yet. The simulations establish that the slowly-driven Floquet systems can exhibit similar relaxation dynamics to the static topological lattices we previously considered in Ref.~\cite{topo_MI}.

\begin{figure}
    \centering
    \includegraphics[width=\columnwidth]{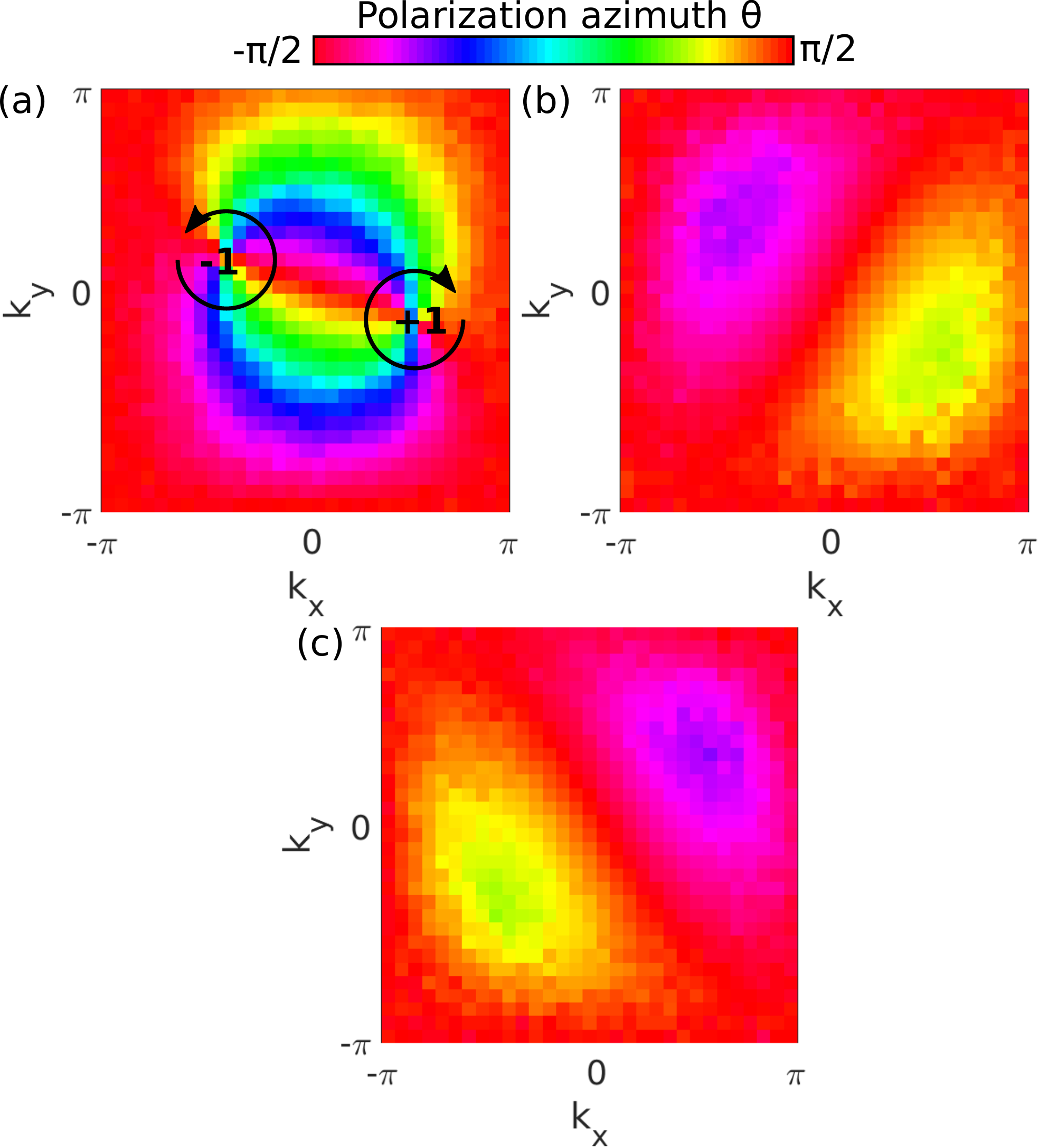}
    \caption{Field polarization profiles in Fourier space, averaged over $100$ random initial perturbations in the three phases. (a) Chern insulator phase ($\gamma = \pi/6$). (b) Anomalous Floquet insulator phase ($\gamma = 7\pi/12$). (c) Anomalous Floquet higher order topological insulator phase ($\gamma = 11\pi/12$).}
    \label{fig:polarization}
\end{figure}

Fig.~\ref{fig:polarization} shows the beam polarization profile in the quasi-steady state (i.e. after $50$ modulation cycles) in the three phases considered.
Vortices in the polarization profile can be created or destroyed in pairs at band crossing points, accompanied by the purity vanishing locally. The weighted sum of polarization vortex charge yields the Chern number. In Fig.~\ref{fig:polarization} we verify that the Chern number is indeed nonzero in the CI phase, while vanishing in the anomalous Floquet phases (since no polarization vortices occur in the quasi-steady state).

\begin{figure}
    \centering
    \includegraphics[width=0.8\columnwidth]{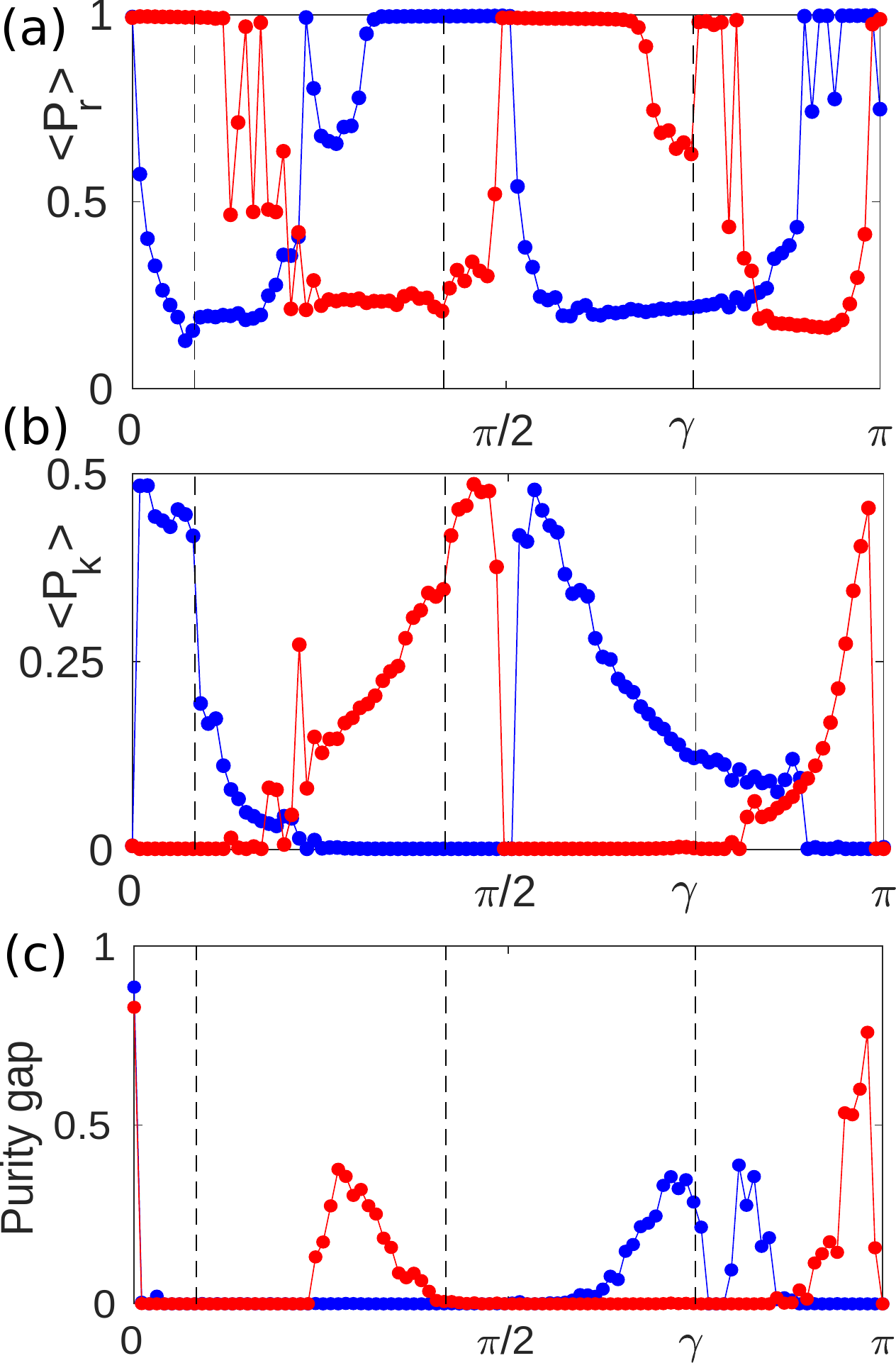}
    \caption{The field measures after $50$ modulation cycles as a function of $\gamma$ for positive (blue) and negative (red) nonlinearity coefficients $|g|=0.35$. Vertical dashed lines indicate the phase transitions. (a) Real space participation number. (b) Fourier space participation number. (c) Purity gap. The purity gap vanishes in the vicinity of the phase transition points and over the range of $\gamma$ where the nonlinear Bloch wave is linearly stable.}
    \label{fig:gamma_scan}
\end{figure}

Finally, Fig.~\ref{fig:gamma_scan} illustrates the beam measures as a function of $\gamma$, indicating the robustness of this quasi-steady-state provided the Floquet band gap remains open and the initial nonlinear Bloch wave is linearly unstable.

In summary, we have seen that the modulational instability of unstable nonlinear Bloch waves enables the complete population of a Floquet quasi-energy band starting from a single wavevector, at least for sufficiently weak nonlinearity strengths. By measuring the polarization of the field after the instability has developed, we can obtain the Bloch wave's $\bs{k}$-dependent polarization profile and thereby the band Chern number, similar to the static lattice case considered previously in Ref.~\cite{topo_MI}. On the other hand, the anomalous Floquet phases, which cannot be distinguished using the Chern number, exhibit singularity-free polarization fields resembling that of a trivial insulator phase.

\section{Measuring anomalous Floquet topological properties}
\label{sec:intra}

Next, we consider the identification of the anomalous Floquet topological phases using the nonlinear propagation dynamics. As noted in Sec.~\ref{sec:model}, the anomalous Floquet topological invariants cannot be obtained by considering the eigenstates of the Floquet evolution operator $\hat{U}(Z)$. Instead, it is necessary to consider the dynamics during the entire driving period, e.g. by considering the phase bands of $\hat{U}(z)$. Unfortunately, the population of a single band of Floquet eigenstates via the modulational instability does not imply population of a single phase band, whose eigenstates can be strongly $z$-dependent. Therefore, one cannot straightforwardly distinguish the anomalous Floquet phases just by measuring the intra-cycle dynamics of the polarization in the steady state.

One might ask whether it is possible to excite only a single phase band by increasing the modulation period $Z$, such that the phase bands undergo an adiabatic modulation. However, the phases we are interested in distinguishing are anomalous Floquet phases, which implies they do not have an adiabatic modulation limit; one inevitably encounters phase band crossing points during the modulation period, meaning that it is not possible to excite only a single phase band. Therefore we need to take a different approach.

Our solution to this conundrum is to exploit the symmetry of the model. First, as was shown in Ref.~\cite{chong}, the anomalous Floquet phases can be distinguished by considering the properties of the phase bands only at the high symmetry points of the Brillouin zone. At these points the phase band eigenstates are independent of $z$; all that changes is the ordering of their eigenvalues, which become degenerate at $z$ values corresponding to band crossing points. At these $z$ values whatever initial polarization state there was at $z=0$ will be restored. By tracking how the polarization rotates in the vicinity of the phase band crossing points, we can extract the Hamiltonian parameters and thereby distinguish the different anomalous phases according to the scheme of Ref.~\cite{chong}.

\begin{figure}[h] 
\center{\includegraphics[width=\columnwidth]{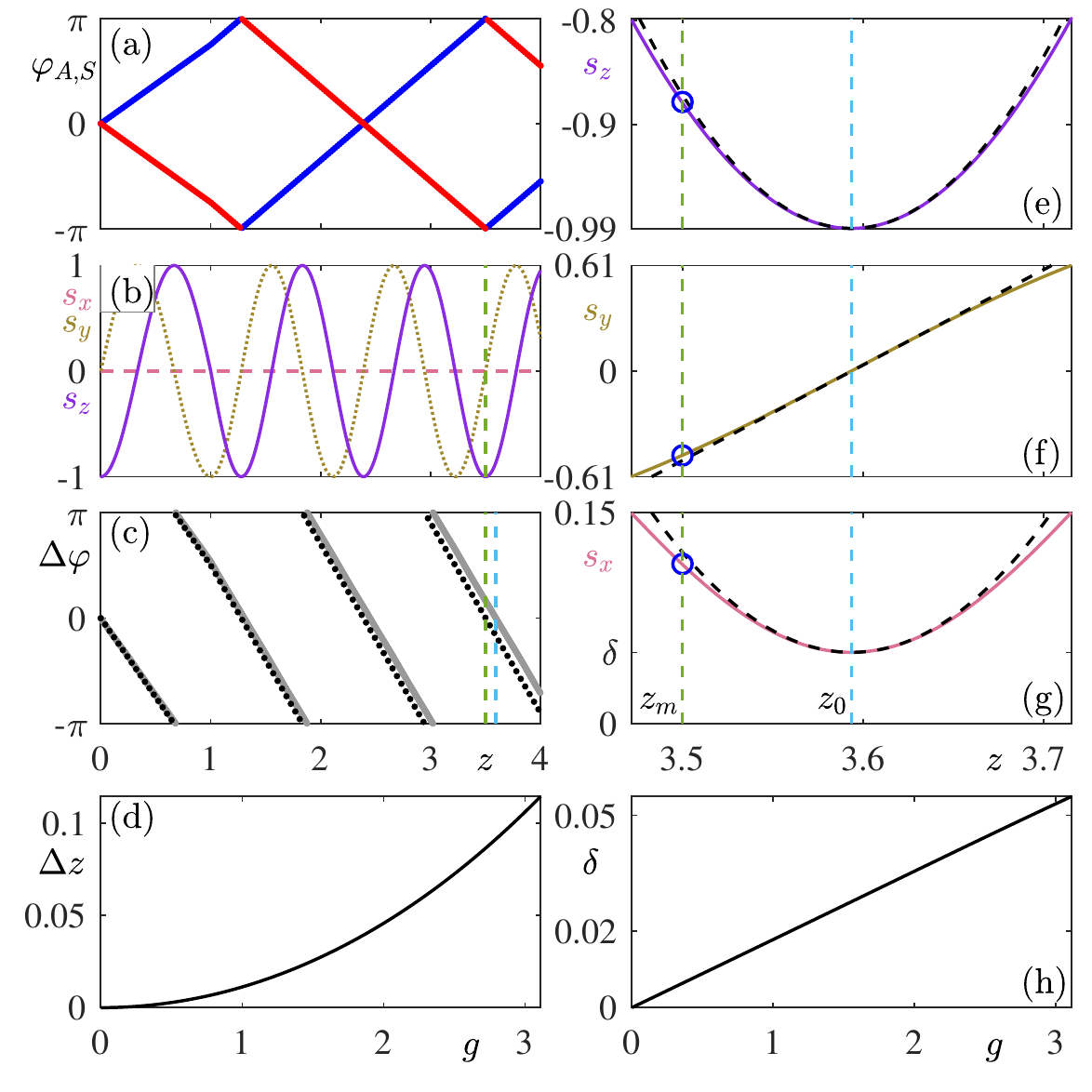}} 
\caption{(a) Phase bands $\varphi_{\text{S,A}}(z)$ at $\Gamma$ point for $\gamma=0.9\pi$, $\theta=3\pi/4$ in the linear regime at $g=0$. (b) Evolution of the spin components in the nonlinear regime over the whole period (solid line $s_z$, dotted line $s_y$, dashed line $s_x$). (c) Phase difference $\Delta \varphi = \varphi_{\text{A}}-\varphi_{\text{S}}$ in the linear case (black dotted line, $g=0$) and $\Delta \varphi = \tan^{-1} {{s_y}/{s_z}}$ in the nonlinear case (gray solid line, $g=0.9\pi$). The vertical green dashed line marks the coordinate $z_m$, at which measurements are taken, $s_x(z_m) = 0.112$, $s_z(z_m) =-0.86587$, $s_y(z_m) =-0.48756$; the vertical cyan dashed line marks the point $z_0$, where $s_y (z_0) = 0$ in the nonlinear case. (e,f,g) Evolution of the spin components in the nonlinear case in the vicinity of $z_0$: blue lines are numerically obtained solutions, dashed black lines are their approximations~\eqref{eq:an_in_vic}. Dependencies of the nonlinear shift $\Delta z = |z_m-z_0|$ (d) and $\delta=s_x(z_0)$ (h) on the nonlinearity strength $g$.} \label{fig3}
\end{figure} 

Figure~\ref{fig3} illustrates the outlined approach for $\Gamma$ point. 
First, we recall that from the dynamic equations for a nonlinear coupler with a coupling strength $J$, 
\begin{equation}
i \partial_z \begin{pmatrix}
a_{\bs{r}}\\
b_{\bs{r}}
\end{pmatrix}  = \begin{pmatrix}
g|a_{\bs{r}}|^2 &  J \\
J  & g|b_{\bs{r}}|^2 
\end{pmatrix} \begin{pmatrix}
a_{\bs{r}}\\
b_{\bs{r}}
\end{pmatrix},
\end{equation}
one can obtain a system of evolution equations for the spin components defined in real space,  $\bs{s}=\bra{\psi_{\bs{r}}}\hat{\bs{ \sigma}}\ket{\psi_{\bs{r}}}$:  
\begin{subequations} \label{eq:spin_evoln}
\begin{align}
\partial_z s_z =  2 J s_y,\\
\partial_z s_y =  - s_z(2J-g s_x),\\
\partial_z s_x =  -g s_y s_z, 
\end{align}
\end{subequations}
and conservation laws for the total intensity and spin
\begin{subequations}
\begin{align}
|a_{\bs{r}}|^2+|b_{\bs{r}}|^2=\mathrm{const} , \\
s_z^2+s_y^2+s_x^2 = \mathrm{const}. \label{eq:spin_consvn}
\end{align}
\end{subequations}
We consider evolution of the initial state $(0,1)^T$, which is a superposition of the symmetric (S) and antisymmetric (A) eigenstates. Figure~\ref{fig3}(a) shows behavior of the phase bands $\varphi_{\text{S,A}}(z)$ over the period $Z$ in the linear case, $g=0$. Assume we gradually cut out slices of the Floquet lattice sample from its end to find empirically, i.e. from polarization measurements, a position $z_m$, where the initial spin is restored $s_z(z_m) = s_z(0) = -1$ and a phase difference vanishes $\Delta \varphi = \varphi_{\text{A}}(z_m)-\varphi_{\text{S}}(z_m)=0$ [see Figs.~\ref{fig3}(b,c)]. Next, we increase the intensity and at this fixed output coordinate $z_m$ measure spin. 

In the nonlinear regime $g \ne 0$, we 
generalize a definition of $\Delta \varphi (z)$ to the function 
$\Delta \varphi = \tan^{-1} ({s_y}/{s_{z}})$, which is plotted with a gray line in Fig.~\ref{fig3}(c) and obeys the equation
\begin{equation}
\partial_z\Delta \varphi = - 2 J + g \dfrac{s_x s_z^2}{s_y^2+s_z^2} = - 2 J + g \dfrac{s_x s_z^2}{1-s_x^2}.
\end{equation}
In Fig.~\ref{fig3}(b) we mark the coordinate $z_0$, where $\Delta \varphi (z=z_0)= 0$ and, therefore, $s_{y} (z_0) = 0 $, whereas $s_{z,x}$ exhibit minima. The total spin conservation law at this point reads
$s_{x0}^2 +  s_{z0}^2 = 1.$
We assume that nonlinearity is relatively weak and the nonlinear shift of $z_0$ with respect to $z_m$ is small. Therefore, at $z_0$ we may approximately write:
\begin{equation}
s_{x0}=\delta,~s_{z0}=-1+\delta^2/2,
\end{equation}   
where $\delta\ll 1$ is a small parameter.
In the vicinity of the point $z_0$, 
we approximate the coordinate dependencies of the spin components as follows: 
\begin{subequations} \label{eq:an_in_vic}
\begin{align}
s_z=-1+\dfrac{\delta^2}{2}+2J^2(z-z_0)^2,\\
s_x=\delta+gJ(z-z_0)^2,\\
s_y=(2J-g\delta)(z-z_0),
\end{align}
\end{subequations}
where $J$ is the (unknown) coupling strength. 
To validate 
expressions~\eqref{eq:an_in_vic}, we substitute Eqs.~\eqref{eq:an_in_vic} into Eqs.~\eqref{eq:spin_evoln} and~\eqref{eq:spin_consvn}, where $|z-z_0| \sim \delta$, and ensure they are consistently fulfilled in the zero $\delta^0$ and first $\delta^1$ orders of perturbation theory. 
In the linear limit, the spin $s_x$ should be precisely equal to zero, thus, we anticipate two asympototics $\delta= G_0 g,~z_m-z_0=G_1g^2$, which are confirmed by  Figs.~\ref{fig3}(d,h).

Taking a measurement of $s_{x\:a} (z_m) \approx G_0 g_a$ at very small but non-negligible intensity $g_a$, the coefficient $G_0 =\tan({{s_{x\:a}}/{g_a}})$ may be determined. Then, perform several measurements with larger nonlinearity $g$ (or equivalently, increasing the intensity of the incident light beam), we can extract $J$ from the system~\eqref{eq:an_in_vic} for each of them and average the results. Thus, the linear dependence of $s_y = 2 J z$ in the vicinity of the phase band crossing points can be used to obtain the coupling strength and distinguish the anomalous Floquet phases from the trivial phase. We note however that, according to Ref.~\cite{chong}, the approach based on the symmetry eigenvalues is still unable to distinguish the AFI phase from the AFHOTI phase.

\section{Conclusion}
\label{sec:conclusion}

Periodically-driven lattices can host a wealth of different topological insulator phases, including Chern insulators, anomalous Floquet insulators, and their higher order analogues. The bulk-edge correspondence originates from the intrinsic relation between the bulk topological properties and the occurrence of topologically protected edge modes. Bulk topological invariants describing anomalous Floquet phases generally depend on the details of the micro-evolution of the system, and have attracted a lot of interest over the past decade. 

However, the observation and measurement  of topological properties in experiments with driven systems is not a fully closed story. Following on our earlier study Ref.~\cite{topo_MI} showing how nonlinear modulational instability in undriven topological lattices can be used to measure their bulk topological invariants, we tested the applicability of this approach to driven systems. The crucial findings of our study can be summarized in a few statements:
\begin{itemize}
\item For weak nonlinearity and gapped Floquet bands, the modulational instability is able to predominantly populate a single band starting from a single Bloch wave, enabling measurement of its Chern number via the polarization profile of the state generated by the modulational instability.
\item To unveil topological properties of anomalous Floquet phases it is generally necessary to consider the microscopic dynamics of the polarization field within the modulation cycle, which is challenging because the microscopic dynamics generally do not involve predominant excitation of a single band of the evolution operator $\hat{U}_L(z)$.
\item In the special case of Floquet lattices with chiral symmetry, anomalous Floquet phases can be distinguished from the trivial phase by studying the time or nonlinearity-dependent dynamics of superpositions of Bloch waves at the high symmetry points of the Brillouin zones.
\end{itemize}
Our methods can be readily implemented using light propagation in nonlinear waveguide arrays, similar to the experiments of Refs.~\cite{Floquet_soliton,edge_soliton_1,edge_soliton_2}, demonstrating the feasibility of using the nonlinear propagation dynamics to measure bulk topological invariants of energy bands.

\section*{Acknowledgements}

This research was supported by the National Research Foundation, Prime Ministers Office, Singapore, the Ministry of Education, Singapore under the Research Centres of Excellence programme, the Polisimulator project co-financed by Greece and the EU Regional Development Fund, the Ministry of Education, Science and Technological Development of the Republic of Serbia (451-03-9/2021-14/ 200017), the Australian Research Council (Grant DE190100430) and the  Institute for Basic Science in Korea (IBS-R024-D1). Numerical analysis of the modulational instability was supported by the Russian Science Foundation (Grant No. 20-72-00148).

\end{document}